\documentclass[12pt]{article}

\usepackage{graphics}
\usepackage{amsmath}
\def\ni{\noindent}
\def\ph{{\phantom{...}}}
\def\={\phantom{..} = \phantom{..}}

\def\+{\phantom{..} + \phantom{..}}
\def\>{\phantom{..} > \phantom{..}}
\def\<{\phantom{..} < \phantom{..}}
\def\-{\phantom{..} - \phantom{..}}

\def\Leq{\phantom{..} \leq  \phantom{..}}

\def\all{\phantom{..} \hbox{for all} \phantom{..}}

\def\bq{\begin{quote}}
\def\eq{\end{quote}}
\def\be{\begin{equation}}
\def\ee{\end{equation}}
\def\bar{\begin{eqnarray}}
\def\ear{\end{eqnarray}}
\def\no{\nonumber}

\def\Sch{Schr{\"o}dinger}

\def\Schist{Schr{\"o}dingerist}
\def\Schists{Schr{\"o}dingerists}

\def\BM{Brownian Motion}

\def\cbE{{\cal E}_{\beta}}

\def\sumr{{\sum_{r=1}^N}}

\def\sumninf{{\sum_{n=1}^{\infty}}}

\def\cE{{\cal E}}
\def\BML{Brownian-Motion-Like}
\def\qps{quasi-periodic signal}

\title{\bf Can \Schist\ Wavefunction Physics Explain \BM? II: The Diffusion Coefficient  \\[2in]
       }

\author{W. David Wick\footnote{email: wdavid.wick@gmail.com}}

\begin{document}
\maketitle
\pagebreak

\section*{Abstract}

In the first paper of this series, I investigated whether a wavefunction model of 
a heavy particle and a collection of light particles might generate ``Brownian-Motion-Like"
trajectories of the heavy particle. I concluded that it was possible, but left unsettled the
second claim in Einstein's classical program: diffusive motion,
proportional to the square-root of time, as opposed to ballistic motion, proportional to the time.
In this paper, I derive a criterion for diffusive motion, 
and an expression for the diffusion coefficient.
Unfortunately, as in paper I, no exact solutions are available for the models, making checking
the criterion difficult.  
But a virtue of the method employed here is that,
given adequate information
about model eigenvalues and eigenfunctions, 
diffusion can be definitively ruled in or out.

\pagebreak

\section{Introduction}

In my first publication on this topic, \cite{WFBMI}, 
I addressed the question of whether \Sch's wavefunction picture of matter
can account for that potent demonstration of the reality of atoms 
from the first decade of the last century: 
Perrin's measurements of the motions of a pollen grain in a water droplet, 
coupled with the theoretical formulas proposed by 
Poincar{\' e} in 1900 and Einstein in 1905. The latter pair explained the irregular motions
of the grain (first observed by Brown in 1828) 
as due to statistical variations in the numbers of water molecules 
colliding with the grain over time. Einstein also linked the diffusion coefficient of the
grain to the temperature and viscosity of the surrounding water bath.

Our challenge today, of course, is that after the revolution of the 1920s, 
we are no longer supposed to believe in classical particles undergoing collisions. 

Paper I includes several simple models of a heavy particle and surrounding light particles,
but as described by wavefunctions, in which the particle's locations are merely arguments
of that wavefunction. It was noted there that the Mean-Square Displacement (MSD) 
of the heavy particle is given by an expression like:

\be
\hbox{MSD} \= \sum_{n=1}^{\infty}\,a_n\,\left[\, 1 - \cos(\nu_n\,t)\,\right]\label{MSDeq}
\ee 

\ni (the $a_n$ are positive coefficients and the $\nu_n$ are frequencies);
in other words, the models produce a ``quasi-periodic signal". However, well-known
models of \BM\ constructed by Wiener, Ornstein, Uhlenbeck and others yield random 
trajectories similar to (continuous-time) ``drunkard's walks" in probability theory.
I argued, starting from Wiener's construction of his process from random Fourier series,
that quasi-periodic signals could in fact yield ``\BML" trajectories
for certain choices of the parameters. 

Also essential to Einstein's
program was a second claim: diffusive behavior, meaning that the MSD grows linearly rather than
quadratically. I noted that, expanding the cosine function in (\ref{MSDeq}), no term
of O($t$) appears, so that, at least for small times, the MSD must grow quadratically.
As our pollen grain cannot escape the droplet under Perrin's microscope slide, the motion
is bounded, so the curve must eventually decline in slope. 
But, I remarked, there might still be an interval of time for which the MSD grows
linearly.

I neglected to note that, depending on whether, e.g.,

\be
 \sum_{n=1}^{\infty}\,a_n\,\nu_n^p < \infty
\ee 
 
\ni for higher powers, this expansion may be uninformative. 
Indeed, as will be shown in the next section, a quasi-periodic signal can grow linearly
at small times. Then I develop a method that
generates proposals for when that is possible or impossible, and a formula for
the diffusion coefficient in the former case. In subsequent sections I apply the method
to the light-particles-plus-heavy particle models. 
Temperature enters into the picture by way of a Gibbs canonical distribution on wavefunctions.

Unfortunately, no formulas are available
for the coefficients in (\ref{MSDeq}), which are derived in the models 
from the eigenfunctions and eigenvalues
of the Hamiltonian (i.e., from a diagonalization). 
In the last section I discuss what proportion of the Poincar{\'e}/Einstein program
the present theory can cover, granted sufficient information about eigenfunctions and eigenvalues.

Several of the theorems stated in this paper are proven {\em gratis} of the computer.
But this helpmate is asked only to produce graphs of two functions of one variable,
given explicitly by simple formulas. I presume any reader of this paper to own a laptop
or other platform, equipped with software implementing graphing routines. It should take that
reader only minutes to check those graphs. Perhaps the word ``proof" nowadays can permit such
minor demands on the reader.

\section{A curious \qps.\label{curioussection}}

Consider the following formula (not attributable to any model, but made up to make a point):

\be
f(t) \= \sumninf\,\left\{\,\frac{1}{n^2}\,\right\}\,\left[\,1 - \cos(2\pi n^2 t)\,\right],
\label{qpform}
\ee

\ni and then look at its graph (generated by adding 1,000 terms at 1,000 time points 
on a computer) in Fig. 1.

\begin{figure}
\rotatebox{0}{\resizebox{5in}{5in}{\includegraphics{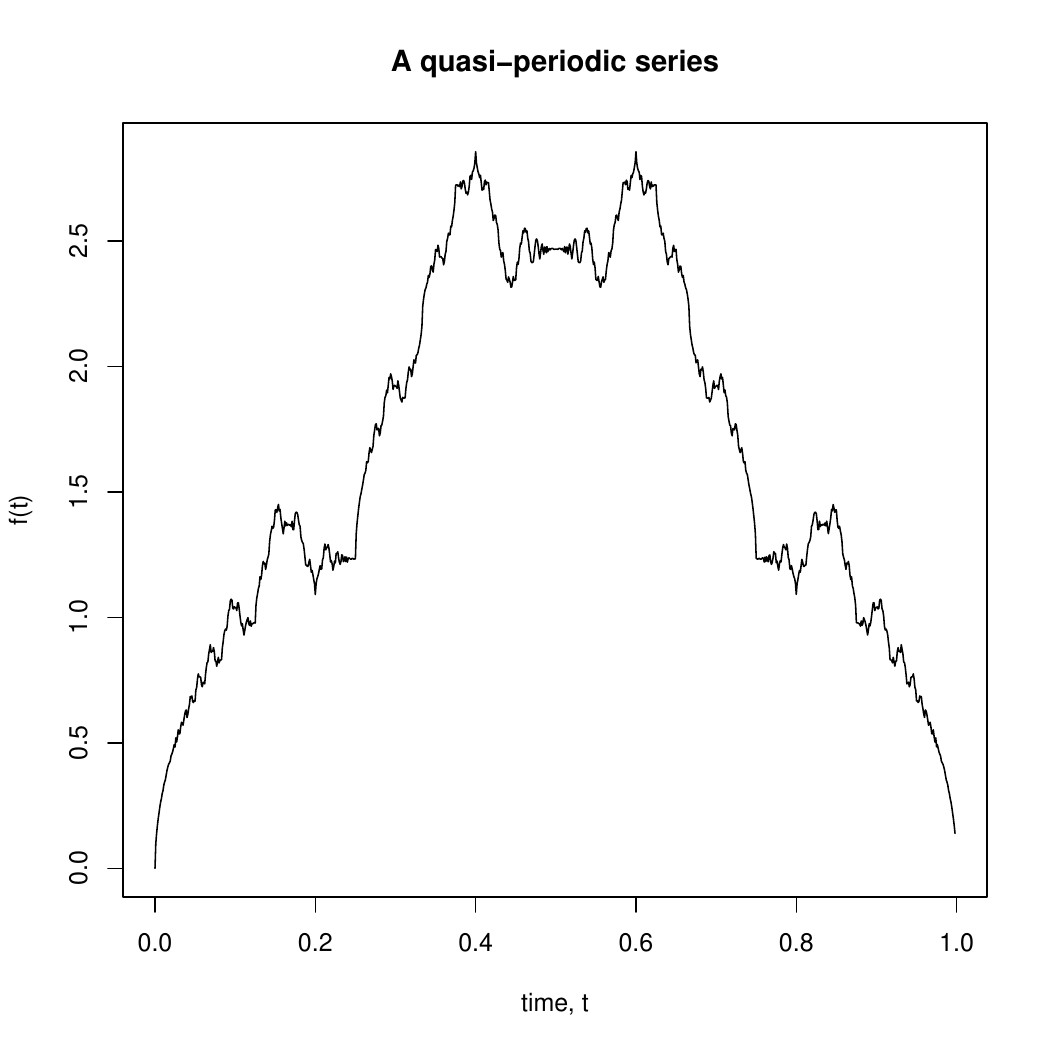}}}
\caption{Graph of the formula (\ref{qpform}) for `$t$' in the unit interval.}
\end{figure}

The curve is left-right symmetric around the midline, $t = 1/2$
(due to the identity: $f(1-t) = f(t)$, a consequence of choosing the frequencies
with the factor of $2\pi$.) But ignore this artifact and examine the left half of the figure;
evidentally, the growth is, granted a little mental smoothing out, linear up until
the midline. 

The latter observation can be rendered into mathematics without arbitrary smoothing.
Fig. 2 shows the left half of the previous graph together with 
the best-fitting quadratic curve.
Note how the latter is essentially linear except for a small negative curvature.

\begin{figure}
\rotatebox{0}{\resizebox{5in}{5in}{\includegraphics{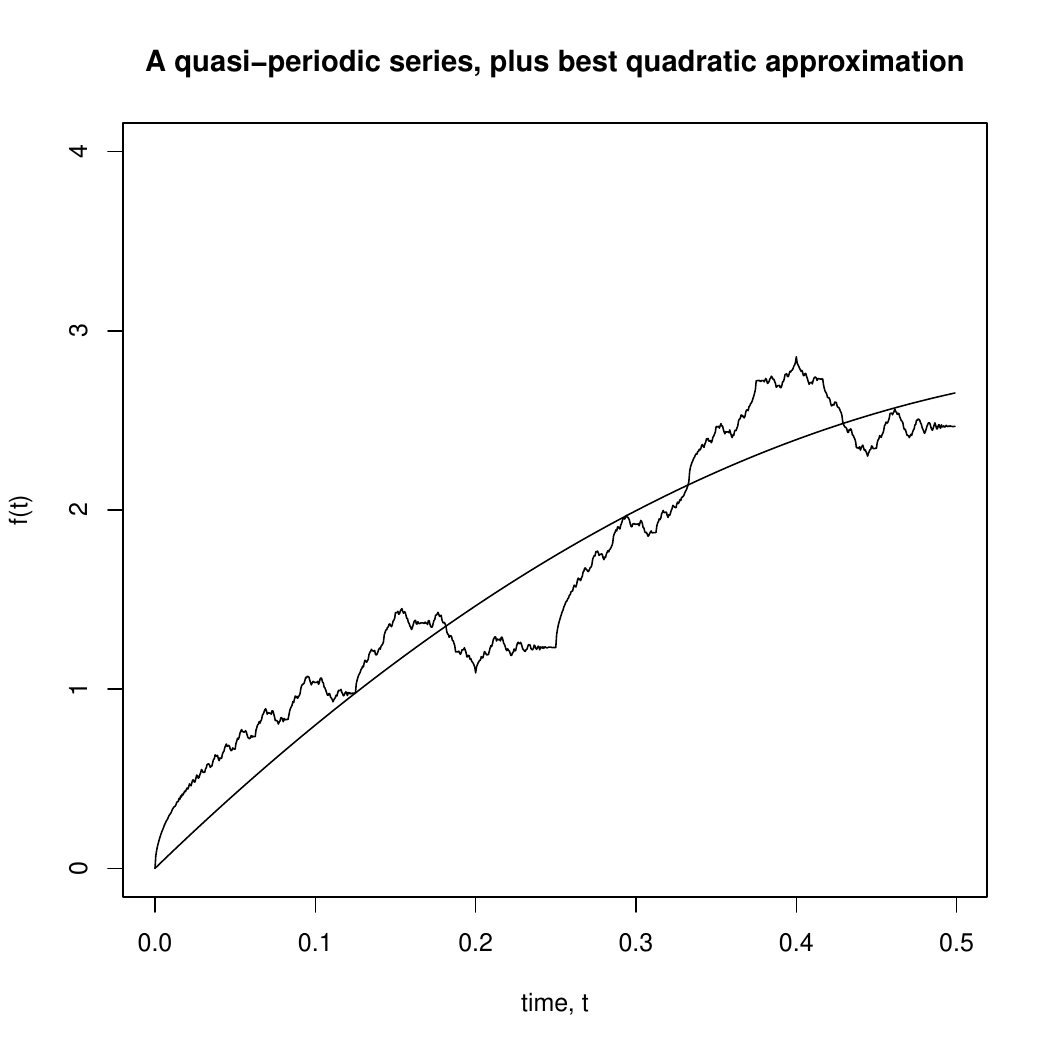}}}
\caption{Left half of the graph of $f(t)$, plus a quadratic curve.}
\end{figure}

We see that expanding the cosine in powers of `$t$' in (\ref{qpform}), 
and concluding that the lowest-order
term contains a positive constant times $t^2$, is misleading, because all of the
resulting sums yield infinite prefactors. Better is to pursue the ``Best Quadratic Fit" (BQF),
a method that is explained 
in detail in the next section.

\section{Theorems about diffusive behavior.}

In the following I assume that $a_n \geq 0$ and $\nu_n > 0$, as will be fulfilled
in applications. (For the latter, looking at (\ref{MSDeq}) reveals that negative frequencies
can be converted to positive, possibly redefining the coefficients $\{a_n\}$.)
I also fix a final time, $T$, which we can interpret as the length of time that Perrin
observed the grain under his microscope. Let $\mu_n = T \nu_n$.

To state the theorems, we require a function of one variable, call it $H(\mu)$, for $\mu > 0$:

\be
H(\mu) \= - \frac{1}{6} - \frac{\sin(\mu)}{\mu} - \frac{5 \cos(\mu)}{\mu^2} - \frac{3 }{\mu^2}
+ \frac{8 \sin(\mu)}{\mu^3}.
\ee

\ni This peculiar definition seems to make for a function with a singularity, and perhaps
a vertical asymptote, at $\mu = 0$, but expanding the sines and cosines in Taylor series
reveals that all singularities cancel out, $H(\mu) \to 0$ as $\mu \to 0$,
 and so $H(\mu)$ extends to the closed interval
$[0,\infty]$ as a continuous function. (In fact, $H(\mu) \approx (1/40) \mu^2 + \cdots$.)
 In the open interval, $H$ is analytic.
The graph of $H$ is shown in Fig 3.  
Note that $H(\mu) \to - 1/6$ as $\mu \to \infty$. 

\begin{figure}
\rotatebox{0}{\resizebox{5in}{5in}{\includegraphics{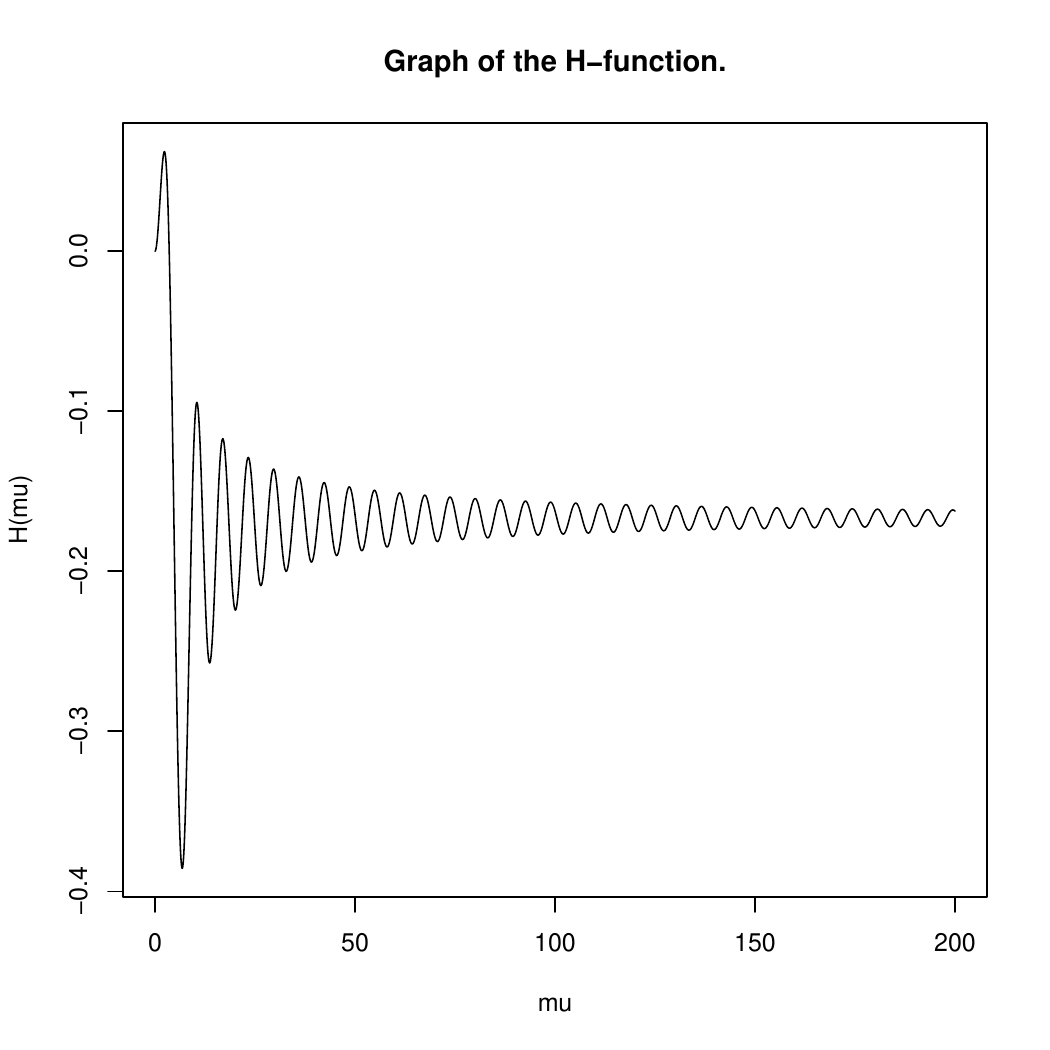}}}
\caption{Graph of the H-function.}
\end{figure}

I now introduce the Best Quadratic Fit (BQF). Let

\be
q(t) \= \alpha\,t + (1/2)\,\gamma\,t^2,\label{quadform}
\ee  

\ni where $\alpha$ and $\gamma$ are real parameters. 
Let $f(t)$ be some real-valued, continuous function on $[0,T]$ with $f(0) = 0$.
Our BQF is that quadratic curve of form given in (\ref{quadform}) that minimizes the
L$^2$-distance:

\be
\int_0^T\,|f(t) - q(t)|^2\,dt.\label{Ldist}
\ee

\ni Let $\alpha^*$ and $\gamma^*$ denote the (unique) minimizing parameters.

\begin{quote} 
{\bf Theorem 1.} Let $f(t)$ be the quasi-periodic function given on the right side of
 (\ref{MSDeq}), and assume $\sumninf a_n < \infty$. Then:

\be
\gamma^* = \left\{\,\frac{40}{T^2}\,\right\}\,\sumninf\,a_n\,H(\mu_n).
\ee

\end{quote}

We take as our criterion for diffusion that:

\begin{quote}
{\bf Diffusion Criterion.}

\be
\gamma^* \leq 0,
\ee

\ni with $\gamma^*$ as given by the formula in Theorem 1. 
\end{quote}

Here is the simplest possibility for diffusion.
Let $\mu_{\hbox{zero}}$ denote the last zero of the function $H(\mu)$ 
on the right-half of the number line.
The computer gave that $\mu_{\hbox{zero}}$ is approximately 3.552.

\begin{quote}
{\bf Theorem 2} 
Suppose that

\be
\mu_n > \mu_{\hbox{zero}} \all n.
\ee

\ni Then the Diffusion Criterion holds.
\end{quote}

\def\HM{H_{\hbox{max}}}

Here is another possibility. 
Let $\HM = \sup\,\{\,H(\mu):0 \leq \mu < \infty\,\}$. Again, the computer gave 
$\HM \approx 0.0621$. 
Given a number $\epsilon$ with $ 0 < \epsilon < 1/6$, let

\be
\mu(\epsilon) \= \inf \,\{\,\mu:\,H(\overline{\mu}) \leq -(1/6) + \epsilon\,\all\,
\overline{\mu} \geq \mu\}.
\ee

From the graph of $H(\mu)$ shown in Fig. 3, $\mu(\epsilon)$ is a decreasing, semi-continuous
function with jumps; $\mu(1/6) = \mu_{\hbox{zero}} > 0$; 
and $\mu(\epsilon) \to \infty$ as $\epsilon \to 0$. 
Let $1[\cdot]$ be the indicator function; i.e., $1[\hbox{condition}] = 1$ 
if the condition is satisfied, and otherwise 0.

\begin{quote}
{\bf Theorem 3.} 

Suppose there exists a number $\epsilon$ with $0 < \epsilon < 1/6$ and

\be
\sumninf\,a_n\,1[\mu_n <= \mu(\epsilon)] \leq \left(\,\frac{1/6 - \epsilon }{\HM}\,\right)\,
\sumninf\,a_n\, 1[\mu_n > \mu(\epsilon)],\label{theorem2ass}
\ee

\ni Then 

\be
\gamma^* \leq 0.
\ee

\end{quote}

Theorem 3 says that a suitable splitting between the lower and the higher frequencies,
together with a bound of the former by the latter, suffices to prove that the 
Diffusion Criterion holds.

For the proofs of Theorems 1 and 3, see the Math Appendix. For how they might be applied 
in models, see Discussion section.

\section{The diffusion coefficient.}

We will need another peculiar function, given by:

\be
G(\mu) \=  \frac{1}{3} + \frac{\sin(\mu)}{\mu} + \frac{6 \cos(\mu)}{\mu^2} + \frac{4 }{\mu^2}
- \frac{10\, \sin(\mu)}{\mu^3}.
\ee

As before, this function is actually non-singular and extends to the closed interval $[0,\infty]$
as a continuous function. It is graphed in Fig. 4. Note that (e.g., from the graph) $G(\mu) \geq 0$
and $G(\mu) \to 1/3 $ as $\mu \to \infty$.

\begin{figure}
\rotatebox{0}{\resizebox{5in}{5in}{\includegraphics{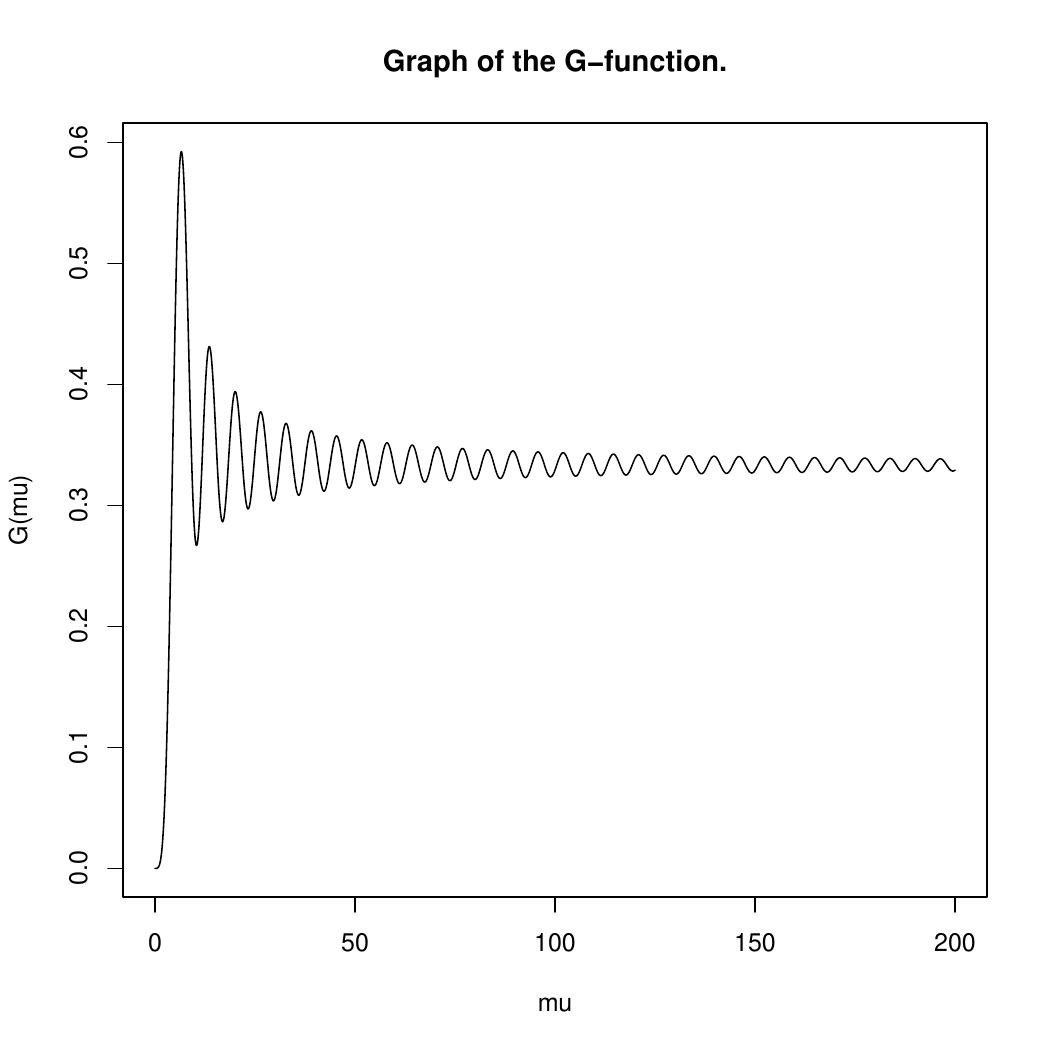}}}
\caption{Graph of the G-function.}
\end{figure}

We can now define a diffusion coefficient by:

\begin{quote} 
{\bf The Diffusion Coefficient} Granted that the Diffusion Criterion holds,
the diffusion coefficient is given by:

\be
  D \= \alpha^*,
\ee

\ni where $\alpha^*$ is the linear coefficient in the BQT to the MSD given in (\ref{MSDeq}).
\end{quote}

\begin{quote}
{\bf Theorem 4.} The diffusion coefficient is given by:

\be
D \= \alpha^* \= \left\{\,\frac{12}{T}\,\right\}\,\sumninf\,a_n\,G(\mu_n).\label{difconstant}
\ee

\end{quote}

That $D$ comes out positive follows from our assumptions.
The proof of Theorem 3 is given in the Math Appendix and properties of the diffusion
coefficient investigated for models in the Discussion section.

\section{Application to the models.}

In paper I several wavefunction models were introduced, which in particle language
might describe a heavy particle (the ``grain") immersed in a bath of light molecules (the
``water molecules", contained in a droplet under Perrin's microscope). The more realistic
model would contain a three-dimensional droplet, with interactions between grain and molecules
given by scattering (repulsive) potentials (ignoring possible excitations of internal states
in either species), and perhaps with pair potentials between water molecules (that might
explain viscosity). But I could not produce a solvable example, meaning such a model in which 
the eigenfunctions and eigenvalues of the Hamiltonian were available in explicit form.  

Therefore, I introduced also a simple, one-space-dimensional, model, in which potentials
were replaced by boundary conditions on the wavefunction prohibiting light particles
on the left of the heavy particle from traversing (tunneling) to the right of it, 
and {\em vice versa}. I called it the ``One Dimensional Toy Model", acronym 1DTM.
I expected that this model could be solved exactly, but was surprised to find it also
intractable (an occurrence that relates to the remarkable fact that the harmonics of
the general right triangle are unknown.) 
With formulas for the eigenfunctions (call them $\psi_n$) and the eigenvalues
($\zeta_n$, yielding frequencies $\omega_n = \zeta_n/\hbar$) unavailable,
I retreated to searching for ``generic" properties of such models.

I let $X$ stand for the heavy particle coordinate and defined the observable to be:

\bar
\no x(t) &\=& <\psi(t)|X|\psi(t)>\\
\no &\=& \sum_{j,k}\,c_k\,c_j^{*}\,<\psi_j|X|\psi_k>\,
\exp\left\{\,i (\omega_k - \omega_j)\,t\,\right\}\\
\no &\=& <\psi(0)|X|\psi(0)> + 
 \sum_{j \neq k}\,c_k\,c_j^{*}\,<\psi_j|X|\psi_k>\,
\left[\,\exp\left\{\,i (\omega_k - \omega_j)\,t\,\right\} - 1\,\right].\\
&&
\ear

There next arose the questions of choosing initial conditions, and how to get temperature
into the game.
\Schists, who do not interpret the wavefunction as a statistical object but rather
as a configuration of matter,
will choose a Gibbsian thermodynamic ensemble of wavefunctions 
(as in De Carlo and Wick,
\cite{decarlowick}, who treated a discrete-spin scenario). 

Identifying each wavefunction in the ensemble with a list of coefficients, e.g.,
$\psi = \sum\,c_k\,\psi_k$, 
the ensemble probabilities may be formally
defined by, 
for any bounded functional of the wavefunction, `$f$': 

\bar
\no \cE_{\beta}\,\left[\,f(c_1,c_2,...,)\,\right] &\=& Z^{-1}\,\int_{\{\sum\,|c_k|^2 = 1\}}\,\prod\,dc_k\,\exp\left\{\,-
\beta\,\sum\,|c_k|^2\,\zeta_k\,\right\}\,f(c_1,...)\\
\no Z &\=& \int_{\{\sum\,|c_k|^2 = 1\}}\,\prod\,dc_k\,\exp\left\{\,-
\beta\,\sum\,|c_k|^2\,\zeta_k\,\right\}.\\
&&
\ear

\ni Here $\beta = 1/(\hbox{Boltzmann's constant} \times\,\hbox{temperature})$.  
To avoid technical problems with defining the integral,
we can restrict the
integrals to a subspace of maximal allowed  
energy, 
e.g., to:

\be
 \sum\,|c_k|^2 =1;\ph \sum\,|c_k|^2\,\zeta_k < E_{\hbox{max.}}.\label{energycondition}
\ee

\ni (One way to do this is to limit the range of `$k$' to be: $1,2,...,N$, for some finite `$N$'.
I.e., assume that no mode above a certain energy ever becomes excited.
This choice will also render all the integrals in this paper 
conventional Riemann, over a $2N$-dimensional sphere.)

In the earlier paper I discussed an equilibrium scenario, in which heavy-plus-light-particles
were initially in thermodynamic equilibrium (when Perrin first saw the pollen grain
through his microscope) and a non-equilibrium scenario. I restrict attention here to the
former scenario.

Next, I defined the mean-squared displacement of the heavy particle averaged over
wavefunctions of the ensemble by:

\be
 \hbox{MSD} \= \cE_{\beta}\,|x(t) - x(0)|^2,
\ee

\ni which in this scenario comes out to be:

\be
 \hbox{MSD} \= 2\,\sum_{j,k: k > j}\,h_{k,j}\,g_{k,j}
\,\left\{\,1 - \cos[(\omega_k - \omega_j)t]\,\right\}.
\label{result}
\ee

\ni Here

\bar
\no g_{k,j} &\=& |<\psi_k|X|\psi_j>|^2;\\
\no h_{k,j} &\=& \cE_{\beta}\,\left(\,|c_k|^2\,|c_j|^2\,\right).\\
&&
\ear

\ni I then relabeled the terms in (\ref{result}) as: 

\be
 \hbox{MSD} \= \sumninf\,a_n\,\left\{\,1 - \cos(\nu_n\,t)\,\right\}.\label{eqformula}
\ee

Such relabeling is always possible, as any countable set can be so ordered (although not uniquely).
A general property of wave- or oscillation-models is that the characteristic frequencies
are increasing: $\omega_j > \omega_k$ for $j > k$, without bound or any cluster point.
This yields the assumption that $\nu_n > 0$, although the sequence $\{\nu_n\}$ may not be ordered
by size and might have zero as a cluster point.

The latter possibility distinguishes these model-derived functions from, e.g., the one presented in
section \ref{curioussection}. Suppose, as an illustration, 
that the frequencies $\omega_k$ are indexed by
N-tuples of integers rather than integers; 
so `$k$' is replaced by: $k \rightarrow (k_1,k_2,...,k_N)$.
Suppose also that these frequencies are given
by:

\be
\omega_k \= \sumr\,b_r\,k_r^2,\label{omegaeq}
\ee

\ni where $(b_1,b_2,...,b_N)$ is an N-tuple of positive real numbers.
(Such a situation appears in the 1DTM. However, formula (\ref{omegaeq})
did not appear in paper I; indeed, I could not obtain formulas for the frequencies of
the 1DTM. However, it did appear in upper and lower bounds on the frequencies of that model.)
Then the `$n$' of our 
observed frequencies $\{\nu_n\}$ becomes a pair of N-tuples: 
$n \rightarrow (k_1,k_2,...,k_N),
(j_1,j_2,...,j_N)$. Hence the $\nu_n$ are given by:

\be
\nu_n \= \sumr\,b_r\,\left(\,k_r^2 - j_r^2\,\right).
\ee

The case of $\nu_n$ zero is ruled out (it makes no contribution to the motion). 
Can the frequency nevertheless be arbitrarily small?
Consider the case $N = 2$, and let $\alpha = b_2/b_1$
be an irrational number.\footnote{That ratios of quantities derived 
from Nature might be given by ratios of integers 
was the fond hope of the Pythagorian School, but is less popular today.} Suppose:

\be 
b_1\,k_1^2 + b_2\,k_2^2 - b_1\,j_1^2 - b_2\,j_2^2 = \epsilon\,b_1.
\ee

If $k_2^2 \neq j_2^2$, re-arranging gives:

\be
\frac{k_1^2 - j_1^2}{j_2^2 - k_2^2} - \alpha \= \frac{\epsilon}{j_2^2 - k_2^2}.
\ee

\ni Since, if $|j_2^2 - k_2^2| \neq 0$ it is at least one, we conclude that:

\be
\big|\,\frac{k_1^2 - j_1^2}{j_2^2 - k_2^2}- \alpha\,\big| \leq \epsilon.
\ee

Is this possible for irrational numbers, for arbitrarily small $\epsilon$? 
Rational numbers are dense in the number line, 
so of course there exist integers $p$ and $q \neq 0$
such that

\be
\big|\,\frac{p}{q}- \alpha\,\big| \leq \epsilon.
\ee

But can each integer be represented as a difference of squares of integers?
Here a little number theory is needed (a self-contained exposition of what we need
is in the Math Appendix). The answer is that an equation of form

\be
p \= k^2 - j^2
\ee

\ni is always solvable for integers $j,k$ provided that $p$ is odd (uniquely if $p$ is prime,
otherwise the number of solutions is finite and can be enumerated). We can assume
both $p$ and $q$ are odd (argued in the Math Appendix). For $N > 2$,
we can produce examples where $j_3 = j_4 = \cdots = k_3 = k_4 = \cdots 1$. 
We conclude that:

\begin{quote}
{\bf Theorem 5}. If, in some model, the frequencies are given by (\ref{omegaeq}), 
then for any $\epsilon >0$
there exists `$n$' such that $\nu_n < \epsilon$. 
Hence the $\{\nu_n\}$ cluster at zero. 
\end{quote} 

The implication of Theorem 5 for checking the diffusion criterion in models
is that, e.g., in the hypothesis of Theorem 3, there will be infinitely-many
terms with low frequencies, appearing on the left side of the inequality, and
ditto for the high frequencies and the right side. Hence checking will not be easy.
However, even lacking exact solutions, we can make some general observations. 

In order to produce either very high or very low frequencies, necessarily the term must
have $k-j$ large. This is obvious for the high frequencies; 
but it is necessary for the low frequencies, too. For example, if a model
had frequencies of form (\ref{omegaeq}), because of the irrationality of $\alpha$,
the `$p$' and `$q$' in the approximation will both have to be large.\footnote{The size 
of integers needed for a rational approximation with error $\epsilon$ of a number
is often taken as a measure of the ``degree of irrationality" of that number.} The size of the
coefficients of such terms will be partially controlled by $g_{k,j}$ in (\ref{result}).
In paper I a bound was established assuming a cut-off on energies in the system, of form: 

\be
\big|<\psi_j|X|\psi_k>\big| \Leq (\hbox{constant})\,\left(\,\frac{1}{|\zeta_j - \zeta_k|}\,\right).
\ee

\ni (See equation (56) and the Math Appendix of that paper.)

The other factor in the term coefficient, $h_{j,k}$, 
will also contribute to suppressing very high and very
low frequencies, because at non-zero temperatures components of the wavefunction with coefficient
$c_k$ will be suppressed for large $k$.

\section{Miscellaneous Comments (mostly about the math).}

The interpretation of Theorem 3 is that, for diffusion to be possible, the lower-
frequency modes must be dominated in total amplitude by around three times the total amplitude
of the higher frequencies.
Is Theorem 3 consistent with a case in which the cosine in (\ref{MSDeq})
can be expanded in a Taylor's series and yields information about small times?
Suppose, for instance, that $\sumninf a_n\,\nu_n^4 < \infty$. Then the Taylor's
approximation to second order and the remainder will be finite when summed over `$n$'.
In the Math Appendix, I show that the assumption in Theorem 3 implies in this situation:

\be
\sumninf \,a_n \leq \left(\,1 + B\,\right)\,\left(\,\sumninf\,a_n\,\nu_n^4\,\right)\,
\left(\frac{T}{\mu(\epsilon)}\right)^4,\label{disceq}
\ee

\ni where $B = (1/6 - \epsilon)/\HM \approx 3$. From this last we can see that
Theorem 3 doesn't apply for small `$T$'. 

On the other hand, what about a model with just one or a few non-zero terms,
and satisfying the hypothesis of Theorem 2? Then the Diffusion criterion holds,
but not the criterion for ``Brownian-Motion-Like" trajectories of paper I.

Do the Theorems 2 and 3 apply to the 
 curious curve of section \ref{curioussection}?
The computer gave that $\mu_{\hbox{zero}}$ is approximately 3.554.
If we choose $T = 0.5$, Theorem 2 does not apply, 
since $2\pi\,T = \pi \approx 3.14$. 

If we write:

\be
S(\epsilon) \= \sumninf\,a_n\,1[\,\mu_n \leq \mu(\epsilon)\,],
\ee

\ni we can rewrite the hypothesis of Theorem 3 as $S(\epsilon) \leq R(\epsilon)$, where:

\bar
\no R(\epsilon) &\=& \frac{B(\epsilon)\,A}{1 + B(\epsilon)};\\
\no B(\epsilon) &\=& \frac{1/6 - \epsilon}{\HM};\\
\no A &\=& \sumninf\,a_n.\\
&&
\ear

At $\epsilon = 1/6$, $S(1/6) = 1$ (since $\pi < 3.55 \approx \mu(1/6) = \mu_{\hbox{zero}}$)
and $R(1/6) = 0$; as $\epsilon \to 0$, $S(\epsilon) \to A$ and $R(\epsilon) \to \approx3\,A/4$.
Hence it not easy, without evaluating more values of $\mu(\epsilon)$ at each jump,
to say whether the hypothesis ever holds.

On the other hand, it is easy to prove diffusion as I have defined it for this curve
directly from the formula for $\gamma^*$. We have $\mu_n = \pi\,n^2$,  
$\sin(\pi\,n^2) = 0$, and $\cos(\pi\,n^2) = \pm 1$, so

\be
\gamma^* \= \frac{40}{T^2}\,\sumninf\,\left\{\,\frac{1}{n^2}\,\right\}\,
\left(\,- \frac{1}{6} +\frac{5\cos(\pi\,n^2)}{\pi\,n^2}
- \frac{3}{\pi^2\,n^4}\,\right),
\ee

\ni which is evidently negative.

Hence, the hypotheses of Theorems 2 and 3, which are sufficient conditions for deducing
diffusion, are not necessary and rather crude.

\section{Discussion: How Much of Einstein's Program Have I Reproduced?}

In this paper I have presented an explicit, checkable, Diffusion Criterion, 
and a formula for the diffusion constant assuming the Criterion holds.
Putting these together with a criterion from paper I for ``Brownian-Motion-Like" 
(BML) trajectories, 
the theory thus far can be summarized as:

\begin{quote}
Suppose that in a heavy-plus-light wavefunction model with positive amplitudes $\{a_n\}$ and
frequencies $\{\nu_n\}$ and observed for a time $T$, you can check that:

\be
\sum_n\,a_n\,\nu_n \= \infty;
\ee

together with either:

\be
\sum_n\,a_n\,H(\nu_n T) \leq 0;
\ee

\ni for a certain universally (not-model-specific) specified function $H$;

Or: the hypothesis of Theorem 2 holds;

Or, the hypothesis of Theorem 3 holds.

Then the model will exhibit BML trajectories and diffusive behavior
with positive diffusion constant given by:

\be
D \= \left\{\,\frac{12}{T}\,\right\}\,\sumninf\,a_n\,G(\nu_n T),
\ee

\ni where $G$ is another universally-specified (and non-negative) function.
\end{quote}

Unfortunately, due to my inability to solve any realistic wavefunction model exactly,
I cannot check the Diffusion Criterion, nor compute the diffusion constant, $D$, as a
function of model parameters and the temperature. Einstein gave the formula:

\be
D \= \frac{K\tau}{6\pi v P},
\ee

\ni where $\tau$ denotes temperature (Einstein used `$T$' but I used it for the
observation time), `$K$' is Boltzmann's constant, 
`$v$' is the viscosity of water (Einstein used `$k$' but we don't want to confuse with 
 a wavefunction index), 
and `$P$'
is the radius of the ``suspended particle" (our ``grain" or heavy particle).

The most interesting number appearing in Einstein's formula is of course, the viscosity.
To even hope to discover 
`$v$' in my formula for $D$ would require a solvable model with intermolecular potentials
plus a wavefunction theory of viscosity. That appears distant. 

It is not even easy to show that
$D$ in my formula increases monotonically with temperature. It is given by:

\be
D \= \cbE\,R;
\ee

\ni where `$R$' is given by:

\be
R \= \left(\,\frac{12}{T}\,\right)\,
\sum_{k,j;k>j}\,g_{k,j}\,|c_k|^2\,|c_j|^2\,G([\omega_k-\omega_j]T).
\ee

\ni Note that $R \geq 0$ and $R=0$ on pure states (eigenstates; $c_{k*}=1$ for some $k*$).
As $\tau \to 0$ ($\beta \to \infty$), $D \to 0$, because the distribution becomes concentrated
on the ground state ($c_1 = 1$). For $\tau$ finite, $D$ will be positive;
as $\tau \to \infty$ ($\beta \to 0$), the distribution reverts to the uniform on the sphere, so
$\cbE\,|c_k|^2|c_j|^2$ is a constant and $D$ tends to a finite or infinite value depending
on whether

\be
\sum_{k,j;k>j}\,g_{k,j}\,G([\omega_k-\omega_j]T)
\ee

\ni is finite or infinite. 

But demonstrating that $D$ is strictly increasing is difficult.
We have that:

\be
\frac{\partial D}{\partial \tau} \= \left(\,\frac{1}{K\,\tau^2}\,\right)\,\left\{\,\cbE\,(RU) 
- \cbE R\,\cbE U\,\right\},
\ee

\ni where $U$ denotes the energy:

\be
U = \sum_k\,\zeta_k \,|c_k|^2.
\ee

$R$ is not functionally increasing with $U$. (Consider a state with two components: 
$c_k = \sin^2(\theta)$ and $c_j = \cos^2(\theta)$, with $k > j$, so $\zeta_k > \zeta_j$,
 and $0 \leq \theta \leq \pi/2$. 
Then $U$ is an increasing function of $\theta$ but $R$ has a local maximum.) 
Owing to the fact that our states are wavefunctions rather than, say, classical spin
configurations, none of the usual theorems of statistical mechanics 
yielding positive correlations apply.
The intriguing possibility that $D$, as a function of temperature, 
has a local maximum remains an open question. 

The reader has surely noted that the scheme presented here is stated entirely in terms of
eigenfunctions and eigenvalues of the Hamiltonian in the linear theory supplied by \Sch.
But \Sch\ never solved the Measurement Problem, which enters here in the question of interpreting
the observable $<\psi(t)|X|\psi(t)>$. For Copenhagenists, 
it is the ``average position of the grain (heavy particle)", 
but not for \Schists; for us, it is the thing observed. However, the statement
``I see that the grain has moved a millimeter to the left" becomes problematic if the wavefunction
is spread out, so that the dispersion in position is of the same order or 
even larger. There is no problem if the wavefunction
is sharply peaked on the left; but in \Sch's linear theory there is nothing to ensure this
scenario. And we cannot accept such facile (even mystical) proposals such as that the wavefunction
``collapses" to the observed position every time Perrin looks into his microscope. This auther
developed, in a series of papers beginning with \cite{WickI}, 
a non-linear generalization of \Sch's theory
which prevents such wave packet spreading in macroscopic components of a measurement apparatus.
(Which raises the question of whether whether Perrin observed such a component. 
As pollen grains are
visible under a store-bought microscope, I would say yes.)

Invoking a nonlinear dynamics will eliminate
the eigenfunction-eigenvalue representation on which the present work is based.
One possibility: with additional information about the model-derived amplitudes and frequencies,
it might turn out that the Diffusion Criterion does not hold. If so, it would be nesessary in 
my program to investigate instabilities (of the kind found in paper III, \cite{WickIII},
of the series)
that might produce apparent ``random" behavior of the grain, which in dynamical 
models is sometimes called ``chaos".

Einstein in 1905 worked in the classical tradition, at least when doing the Brownian Motion theory.
Any mathematician desiring to make a rigorous model of the motion as he described it
would certainly end up with a stochastic process, as did Wiener, Ornstein and Uhlenbeck.
But such models had to contend, in the first decade of that century, with claims that
what Perrin had observed was really some kind of oscillation. The theory presented here seems
to revert to that rejected picture, but perhaps brings some clarity to the dispute.
I would agree that wavefunction theory produces an ``oscillation" provided only
a few frequencies contribute to the observed motion, or if the series of sines or cosines
and its time-derivative converged absolutely. But, if it turns out that the BML and Diffusion
critera can be met, the series that passes the test will be more like Wiener's construction
of his stochastic process from Fourier series than like what appears in the theory of a 
simple pendulum.

\section*{Math Appendix}

\subsection*{Proofs of the Theorems}

To prove Theorems 1 and 4
 we need to minimize the L$^2$-distance given in (\ref{Ldist}) with respect to
the parameters $\alpha$ and $\gamma$. That distance is a convex function of the parameters,
hence the minimum occurs at a critical point for which both partial derivatives vanish.
For ease of writing these equations let:

\bar
\no I_1 &\=& \int_0^T \, t\,f(t)\,dt;\\
\no I_2 &\=& \int_0^T \, t^2\,f(t)\,dt.\\
&&
\ear

The conditions of vanishing partial derivatives then yield the pair of equations:

\bar
\no 8\alpha + 3T\gamma &\=& \left(\,\frac{24}{T^3}\,\right)\,I_1;\\
\no 5\alpha + 2T\gamma &\=& \left(\,\frac{20}{T^4}\,\right)\,I_2;\\
&&
\ear

\ni which in matrix form is:

\be
\begin{pmatrix}
8 & 3T\\
5 & 2T
\end{pmatrix}\,
\begin{pmatrix}
\alpha\\
 \gamma
\end{pmatrix} \=
\begin{pmatrix}
24\,I_1/T^3\\
20\,I_2/T^4
\end{pmatrix}
\ee

The determinant of the square matrix on the left side is $T>0$, so inverting:

\be
\begin{pmatrix}
\alpha \\
\gamma
\end{pmatrix} \=
\begin{pmatrix}
1 \\
T
\end{pmatrix}
\begin{pmatrix}
2T & -3T\\
-5 & 8
\end{pmatrix}\,
\begin{pmatrix}
24I_1/T^3\\
20I_2/T^4
\end{pmatrix}
\ee

Multiplying out we obtain the equations for the critical parameters:

\bar
\no \alpha^* &\=& \left(\,\frac{12}{T^4}\,\right)\,\left(\, 4TI_1 - 5I_2\,\right);\\
\no \gamma^* &\=& \left(\,\frac{40}{T^5}\,\right)\,\left(\, -3TI_1 + 4I_2\,\right).\\
&&\label{criteqns}
\ear

Now let $f(t)$ be given by (\ref{MSDeq}). Plugging into the definitions of $I_1$ and $I_2$,
the resulting integrals are elementary (the ones involving powers of `$t$' multiplying
trig functions are performed by several integrations-by-parts). The results are:

\bar
\no I_1 &\=& \sumninf\,a_n\,\left\{\, \frac{T^2}{2} - \right(\,\frac{T}{\nu_n}\,\left)\sin(\nu_n T)
- \frac{1}{\nu_n^2}\,\left[\,\cos(\nu_n T) - 1\,\right]\,\right\};\\
\no I_2 &\=& \sumninf\,a_n\,\left\{\, \frac{T^3}{3} - \right(\,\frac{T^2}{\nu_n}\left)\,
\sin(\nu_n T)
- \left(\,\frac{2T}{\nu_n^2}\,\right)\,\cos(\nu_n T)
  + \left(\,\frac{2}{\nu_n^3}\,\right)
\,\sin(\nu_n T)\,\right\}.\\
&&\label{Ieqns}
\ear

The expressions in (\ref{Ieqns}) can now be substituted into (\ref{criteqns}) and the latter
equations simplified; this yields Theorems 1 and 4.

Theorem 2 follows immediately granted that Fig.3 is correct.  

The proof of Theorem 3 is simple: from the definition of $\mu(\epsilon)$ and assuming $a_n >= 0$:

\bar
\no \sumninf\,a_n H(\mu_n) &\=& \sumninf\,a_n\,H(\mu_n)\,1[\, \mu_n \leq \mu(\epsilon)\,] \+
\sumninf\,a_n\,H(\mu_n)\,1[\, \mu_n > \mu(\epsilon)\,]\\
\no &\leq& \HM\,\sumninf\,a_n \,1[\, \mu_n \leq \mu(\epsilon)\,] \+
\left(\,-1/6 + \epsilon\,\right)\,\sumninf\,a_n\,1[\, \mu_n > \mu(\epsilon)\,]\\
&&
\ear

\ni from which the theorem follows.

\subsection*{Proof of (\ref{disceq}) in Misc. Comments}

We can write:

\be
\left(\frac{1}{A}\right)\,\sumninf\,a_n\,1[\mu_n > \mu_{\epsilon}] \=
P\left[\,\mu > \mu_{\epsilon}\,\right],
\ee

\ni where we think of $\mu$ as a random variable taking value $\mu_n$ with probability
$a_n/A$. Jensen's inequality then gives:

\be
P\left[\,\mu > \mu_{\epsilon}\,\right] \leq \left(\frac{1}{\mu_{\epsilon}}\right)^4\,\hbox{E}
\,\mu^4,
\ee

\ni where `E' in this inequality denotes expectation. If we define:

\be
U_{\epsilon} \= \sumninf\,a_n\,1[\mu_n > \mu_{\epsilon}],
\ee

\ni then the assumption of Theorem 3, given in (\ref{theorem2ass}), reads

\bar
\no A - U_{\epsilon} &\leq& B(\epsilon)\,U_{\epsilon};\\
\no A &\leq& (1+B(\epsilon))\,U_{\epsilon},\\
&&
\ear

\ni which putting all together yields
(\ref{disceq}).

\subsection*{A little number theory}

Given a positive integer `$p$', let $P(p)$ denote the number of distinct ways of representing
$p$ as a product: $p = q_2\,q_1$, with $q_1$ and $q_2$ positive integers (unity is allowed) 
and $q_2 > q_1$.

\begin{quote}
{\bf Difference-of-Squares Lemma}
Let `$p$' be a positive, odd integer. Than the equation:

\be
p \= k^2 - j^2
\ee

\ni has the solution in positive integers:

\be
k = \frac{p+1}{2};\ph j = \frac{p-1}{2}.
\ee

\ni If $p$ is prime, this solution is unique. Otherwise,
there are $P(p)$ distinct solution pairs.
\end{quote} 

For example, $7 = 4^2 - 3^2$, which is unique, while $15 = 8^2 - 7^2$ and also $4^2 - 1^2$.
But 6 is not a difference-of-squares.

{\bf Proof of the DSL}: 

Let $p = q_1\,q_2$ with $q_2 > q_1$ and both odd; setting:

\be
k = \frac{q_2 + q_1}{2}; j = \frac{q_2 - q_1}{2};\label{kjeq}
\ee

\ni yields a solution. Conversily, given a solution pair $(j,k)$, factoring $p$:

\be
p = (k - j)\,(k + j),
\ee

\ni it must be the case that: 

\be
k - j = q_1; k + j = q_2;
\ee

\ni for some decomposition of $p$; from which (\ref{kjeq}) follows. QED.

{\bf The remark that $q$ and $p$ can be taken odd:} The issue is whether
we can approximate:

\be
\big|\,\frac{p}{q} - \alpha\,\big| < \epsilon,
\ee

\ni with both $p$ and $q$ odd. If we have such an approximation but $q$ is even,
we can certainly obtain a better one substituting $q+1$ for $q$. So we can assume
$q$ is odd. Now suppose $p$ is even and 

\be
\big|\,\frac{p}{q} - \alpha\,\big| < \frac{\epsilon}{2}; \frac{1}{q} < \frac{\epsilon}{2}; 
\ee

\ni then 

\be
\big|\,\frac{p+1}{q} - \alpha\,\big| \=   
\big|\,\frac{p}{q} + \frac{1}{q} - \alpha\,\big| \leq \epsilon,   
\ee

\ni proving the remark.


\begin{thebibliography}{9}

\bibitem{WFBMI}
Wick, W. D. Can \Schist\ Wavefunction Physics Explain \BM? Arxiv quant-ph 2305.11977.
19 May 2023.


\bibitem{einstein} {\em Investigations on the Theory of the Brownian Movement}.
Dover, NY, NY. 1959. Reprints of Einsteins's papers on \BM, translated into English by 
A. D. Cowper.

\bibitem{perrin} Perrin, J, {\em Atoms}. Translated by D. L. Hammick. 1916.
Constable \& Co, Limited. London.

\bibitem{decarlowick}
De Carlo, L. and Wick, W.D. ``On \Schist\ Quantum Thermodynamics". 16 August 2022.
ArXiv 2208.07688. Journal publication: ``On Magnetic Models in Wavefunction Ensembles."
{\em Entropy} 25(4) 564 (2023).

\bibitem{WickI} 
Wick, W.D. ``On Non-linear Quantum Mechanics and the Measurement Problem I. Blocking Cats".
ArXiv 1710.03278 (2017).

\bibitem{WickIII} 
Wick, W.D. ``On Non-linear Quantum Mechanics and the Measurement Problem 
III: Poincar{\'e} Probability and ... Chaos?".
ArXiv 1803.11236 (2018).

\end{thebibliography}
\end{document}